\documentclass[11pt,preprint]{aastex}
\def\feh{{\rm [Fe/H]}}

\def\Teff{T_{\rm eff}}

\def\popi{Population~I}
\def\popii{Population~II}

\def\Str{Str\"omgren}
\def\subsun{_\odot}

\begin{document}

\title{The \emph{uBVI} Photometric System. I. Motivation, Implementation, and
Calibration}

\author{Howard E.\ Bond\altaffilmark{1}}

\affil{Space Telescope Science Institute, 3700 San Martin Drive, Baltimore, MD
21218; bond@stsci.edu}

\altaffiltext{1}
{Visiting Astronomer, Kitt Peak National Observatory and Cerro Tololo
Interamerican Observatory, National Optical Astronomy Observatory, which are
operated by the Association of Universities for Research in Astronomy, Inc.,
under cooperative agreement with the National Science Foundation.}

\begin{abstract}
This paper describes the design principles for a CCD-based photometric system
that is highly optimized for ground-based measurement of the size of the Balmer
jump in stellar energy distributions. It is shown that, among ultraviolet
filters in common use, the Thuan-Gunn $u$ filter is the most efficient for this
purpose. This filter is combined with the standard Johnson-Kron-Cousins $B$,
$V$, and $I$ bandpasses to constitute the \textit{uBVI} photometric system. 

Model stellar atmospheres are used to calibrate color-color diagrams for the
\textit{uBVI} system in terms of the fundamental stellar parameters of effective
temperature, surface gravity, and metallicity.  The $u-B$ index is very
sensitive to $\log g$, but also to $\feh$\null. It is shown that an analog of
the \Str\ $c_1$ index, defined as $(u-B)-(B-V)$, is much less metallicity
dependent, but still sensitive to $\log g$.  The effect of interstellar
reddening on $u-B$ is determined through synthetic photometric calculations, and
practical advice is given on dealing with flat fields, atmospheric extinction,
the red leak in the $u$ filter, and photometric reductions.

The \textit{uBVI} system offers a wide range of applicability in detecting stars of
high luminosity in both young (yellow supergiants) and old (post-AGB stars)
populations, using stars of both types as standard candles to measure
extragalactic distances with high efficiency, and in exploring the horizontal
branch in globular clusters. In many stellar applications, it can profitably
replace the classical \textit{uBVI} system.

Paper~II in this series will present a network of well-calibrated standard
stars for the \textit{uBVI} system.
\end{abstract}


\section{Introduction and Motivation}

Stars of high intrinsic brightness are essential in a variety of astronomical
applications, including their use as standard candles for measuring
extragalactic distances, and as luminous tracers of stellar populations in
external galaxies. In the future, such objects will serve as dynamical and
chemical probes, through measurements of radial velocities and determinations
of chemical abundances with spectrographs on large telescopes, and through
space-based microarcsecond astrometry. 

The visually brightest non-transient stars are \popi\ supergiants of A, F, and
early~G spectral types, hereafter called ``yellow supergiants.'' The rarest and
most luminous of these stars attain visual absolute magnitudes as bright as
$M_V\simeq-10$ (Humphreys 1983).  

Yellow supergiants are the optically brightest stars because of the behavior of
the bolometric correction, which is smallest at these spectral types but becomes
quite large for blue and red supergiants.  Fig.~1
illustrates this point. On the left are plotted evolutionary tracks for massive
stars (Meynet et al.\ 1994), in the usual theoretical coordinates of $\log
L/L_\odot$ vs.\ $\log T_{\rm eff}$\null. The plotted tracks are for solar
metallicity ($Z=0.02$ by mass), and the models include mass loss. Each track is
labelled with the star's initial main-sequence mass. Following an initial rise
off the main sequence, the stars evolve to cooler effective temperatures at
roughly constant luminosity (with the more massive stars eventually turning
back toward higher temperatures near the ends of their lives, and the less
massive rising to very high luminosities as red supergiants).

On the right side of Fig.~1, these tracks have been converted to the
observational quantities $M_V$ vs.\ $B-V$, using the formulae of Flower (1996).
Now we see the dramatic fashion in which the observational tracks peak in
brightness around types~A and F (i.e., between $B-V\simeq0$ and 0.5). The sharp
peaking is due to the strong dependence of the bolometric correction upon
$T_{\rm eff}$, which overwhelms the evolutionary variations in bolometric
luminosity. 

Likewise, the brightest stars of \popii\ are of spectral types A and F, but in
this case they are low-mass post-asymptotic-giant-branch (PAGB) stars evolving
off the tip of the AGB and passing through these spectral types on their way to
the top of the white-dwarf cooling sequence.   PAGB stars of these spectral
types show great promise as \popii\ standard candles and as luminous tracers of
old populations (Bond 1997; Alves, Bond, \& Onken 2001).

In addition to their high optical luminosities, yellow supergiants and PAGB
stars have the advantage of being easily detected through multicolor
photometry. The reason is that, because of their very low surface gravities,
their spectral energy distributions show extremely large Balmer discontinuities
in the optical UV region.  No other stellar objects have such large Balmer
jumps, giving these stars a unique photometric signature that can be detected
at low spectral resolution even at faint apparent magnitudes.  Moreover, this
signature can be detected in a single observation, unlike the time-series
photometry needed to detect the subset of cooler and fainter yellow supergiants
that are Cepheid variables.

The problem, however, is that historically there has not been a widely used
photometric system that is fully optimized for the detection of faint stars
with large Balmer jumps.  The classical Johnson-Kron-Cousins $UBVRI$ system has
been in use for nearly five decades, and extensive calibrations and
networks of standard stars have been established through the heroic work of
Landolt (1992) and many others. However, the $U$ filter of this system
straddles the Balmer jump, and is thus far from optimum for measuring its size.
(In fact, it is highly undesirable to straddle the jump, since when the size of
the discontinuity increases due to lowering of the surface gravity, much of the
extra absorbed flux is redistributed to wavelengths just longward of the jump.)
By contrast, the $u$ filter of the $uvby$ system (\Str\ 1963) was specifically
optimized to measure the Balmer jump by placing all of its transmission below
$\sim$3650~\AA; however, all four filters of this system have
intermediate-width bandpasses, making the throughput undesirably low for
efficient observations of faint stars in external galaxies. The
intermediate-band $uvgr$ system introduced by Thuan \& Gunn (1976) has the
advantage of a $u$ filter lying almost entirely below the Balmer jump, but did
not, at the time the work reported here was planned, have the wide usage and
extensive calibration work of the $UBVRI$ and \Str\ systems for stellar
photometry, especially in the southern hemisphere. The Thuan-Gunn system was
modified to the wide-band $u'g'r'i'z'$ filters for the Sloan Digital Sky Survey
(Fukugita et al.\ 1996), but unfortunately (in the present context), in order
to increase its throughput for work on faint extragalactic objects, the
designers of this system adopted a $u'$ filter with significant transmission
above the Balmer discontinuity.

The purpose of the present paper is to describe a new CCD-based ``\textit{uBVI}''
photometric system, which is highly optimized for detection of faint stars with
large Balmer jumps, has high throughput for all four filters, and whose $BVI$
filters can be readily calibrated against the large body of existing stellar
photometry that has accumulated over the past decades. Because of the wide use
of the $UBVRI$ system for stellar work, including its extensive use for
measurements of extragalactic standard candles, and its broad bandpasses, the
writer quickly decided that, above the Balmer jump, one should simply adopt the
standard $B$, $V$, and $I$ filters of the Johnson-Kron-Cousins
system.\footnote{Similar considerations led Kinman et~al.\ (1994) to adopt and
use a $uBV$ system for photoelectric photometry of A stars in the galactic
halo, where the $u$ filter is that of the \Str\ system; however, as noted
below, the low throughput of \Str\ $u$ makes it non-optimal for faint stars.} 
The $I$ band was included because of its low sensitivity to interstellar
extinction, and because  $V-I$ provides a color index less sensitive to
metallicity than $B-V$; moreover, the $I$ measurement provides an elegant means
for dealing with the red leak of the chosen $u$ filter (see below). (For the
sake of efficiency, it was felt that the $R$ filter would not provide
sufficient additional information to justify its addition to the system.)

The following sections describe the selection of a UV filter to be added to
$B$, $V$, and $I$; the calibration of the \textit{uBVI} system based on model stellar
atmospheres; determination of the interstellar extinction coefficients; and a
discussion of some practical considerations in implementing this new
photometric system.  Paper~II in this series (Siegel \& Bond 2005) will
describe our establishment of a network of equatorial standard stars for our
\textit{uBVI} system.

\section{Selection of an Optimal UV filter}

What is the optimum choice for a UV filter to measure the size of the Balmer
jump?  A narrow-bandpass filter lying entirely below the jump would have high
sensitivity to the size of the discontinuity, but low observing efficiency.  At
the other extreme, a wide filter extending somewhat above the jump would
transmit more photons, but would be less sensitive to the size of the
discontinuity, especially because of the flux-redistribution effect described
above.  In typical applications, about 3/4 of the total observing time goes
into the $u$-band exposures, so it is crucial to choose the most efficient
filter if one plans to observe faint stars.

To make this efficiency assessment, I calculated a ``figure of merit'' for
measuring the Balmer jump as follows.  I consider an idealized experiment in
which two solar-metallicity stars of the same apparent angular radius, both
having $\Teff=7000$~K (near the effective temperature at which the size of the
jump is largest), are observed separately, one of surface gravity $\log g=4.5$
(main-sequence star) and the other with $\log g=1.0$ (typical of a \popi\
yellow supergiant or \popii\ PAGB star).  Both stars are to be observed through
the $u$ and $B$ filters, and the change in the instrumental $u-B$ color index
is the measure of the change in the Balmer jump. The figure of merit then
becomes the total exposure time (summed over the $u$ and $B$ observations of
both stars) needed to measure the change in $u-B$ color index, $\Delta(u-B)$,
to a given accuracy, say 1\% of $\Delta(u-B)$. The optimum observing procedure
would be to spread the error budget evenly over all 4 exposures, i.e., detect
the same number of photons in each of the 4 exposures. In this idealization, I
neglect such factors as sky background, and simply take the measurement errors
to be proportional to the inverse square root of the number of detected
photons.

Calculations were made using the computational technique and model stellar
atmospheres described in detail below, assuming unreddened stars and an airmass
of 1.2. The detector sensitivity of the Cerro Tololo 0.9-m CCD camera, also
described below, was adopted.  The actual Johnson $B$ filter used in this
camera was assumed, and four different candidate UV filters were considered:
Thuan-Gunn $u$, \Str\ $u$, SDSS $u'$, and Johnson $U$\null. Figure~2 plots the
transmission curves of these four filters. In order to show the location of the
Balmer jump,  the flux distribution of a model atmosphere with $\Teff=7000$~K,
$\log g=1$, and $\feh=0$, taken from Lejeune, Cuisinier, \& Buser (1997,
hereafter LCB97), is also plotted. Results of the calculations are presented in
Table~1. The notes at the end of Table~1 give the sources for the filter
transmission curves used in the calculations.

The table shows that the Thuan-Gunn $u$ filter is the best choice by a
considerable margin. Although \Str\ $u$ is more {\it sensitive\/} to surface
gravity [that is, it has the largest $\Delta(u-B)$, because it transmits
essentially no flux above 3650~\AA], it is not the most {\it efficient\/},
because of its relatively low throughput. The $u'$ and $U$ filters have higher
throughput, but lower sensitivity to $\log g$ because they have significant
transmission above the Balmer jump, and are thus likewise
sub-optimal.\footnote{It should be noted that Table~1 somewhat overstates the
figures of merit for SDSS $u'$ and Johnson $U$, since actually measuring the
Balmer jumps in the test stars to 1\% accuracy would require measuring the
color indices to accuracies of 0.0049 and 0.0038~mag, respectively.  At
accuracy levels this demanding, various sources of systematic errors 
(short-term variations in atmospheric transmission,  flat fields,
transformation errors, etc.)\ typically start to become important compared to
simple photon statistics. By contrast, the more modest requirement of
$\sim$0.0085~mag accuracy for Thuan-Gunn or \Str\ $u$ is generally easier to
achieve.}

I therefore chose to adopt the Thuan-Gunn $u$ as the ultraviolet filter.
Details of the Thuan-Gunn $u$ filter (hereafter called simply ``$u$'' when
there is no ambiguity), including the recipe for constructing it, and a
measured transmission curve, are given in Appendix~A. 

Figure~3 plots the transmission curves for the \textit{uBVI} filters alone, and
the total system throughput curves, for the Cerro Tololo Interamerican
Observatory (CTIO) 0.9-m telescope, filters, and
CCD camera, viewing through an airmass of 1.2 (calculated as described below).
Also shown, again in this figure, is the LCB97 flux distribution for
a low-gravity F star ($\Teff=7000$~K, $\log g=1$, $\feh=0$).

Table~2 summarizes some basic properties of the four filters of this new \textit{uBVI}
system. The final column lists references for the nominal effective wavelengths
and full widths at half maximum (FWHM) for the filters.  As noted in the table
and discussed below, the zero points for the \textit{uBVI} magnitudes will be set such
that Vega will have magnitude zero in $BVI$, but magnitude 1.00 at $u$.

\section{Calibration of Stellar Atmospheric Parameters}

\subsection{Computational Method}

In this section I present calculations of the dependence of colors in the
\textit{uBVI} system upon the fundamental stellar parameters of effective temperature,
surface gravity, and metallicity.

Stellar magnitudes, $m$, are defined with the usual equation
\begin{equation}
m =  -2.5 \log n_{\rm phot} + {\rm const} \, , 
\end{equation}
where $n_{\rm phot}$ is the number of photons detected from a star using a
given filter, CCD camera, and telescope system.  The number of detected photons
is given by the following equation:
\begin{equation}
n_{\rm phot}\propto \int_{0}^{\infty} 
\lambda F_\lambda(\Teff,\log g,\feh; \lambda) \,
I[E(B-V), \lambda] \,
A(\lambda)^X \,
r(\lambda)^2 \,
T(\lambda) \,
Q(\lambda) \,
d\lambda \, .
\end{equation}

The terms in eq.~2 have been arranged in order from the star to the
detector, and have the following meanings:

1. $\lambda F_\lambda(\Teff,\log g,\feh;\lambda)$ is the photon flux from a
star of a given effective temperature, surface gravity, and metallicity. 
$F_\lambda$ is multiplied by the wavelength $\lambda$ in eq.~2 in order to
convert energy flux to photon flux, since the CCD detector signal is
proportional to the number of photons. Throughout this paper I have obtained
stellar fluxes from the library of ``corrected'' synthetic stellar spectra
presented by LCB97. LCB97 actually tabulate the Eddington flux $H_\nu$, which,
as they note, should be converted to $F_\lambda$ using $F_\lambda=0.4 H_\nu
c/\lambda^2$.

2. $I[E(B-V), \lambda]$ is the attenuation due to interstellar extinction. I
have used the analytic formulae of Cardelli, Clayton, \& Mathis (1989) with
$R_V$ set to 3.1.

3. $A(\lambda)$ is the Earth's atmospheric transmission as a function of
wavelength at an airmass $X=1$.  This function was set equal to  the mean of
the atmospheric extinction curves measured on 38 nights of spectrophotometry at
CTIO between 1988 and 1994 by Hamuy et~al.\ (1994), kindly provided to the
writer in tabular form by Hamuy (1999). I make the assumption that the same
curve can be used to simulate observations made at Kitt Peak National
Observatory (KPNO)\null.  Hamuy tabulates
extinction coefficients in magnitudes per airmass, $k(\lambda)$, which are
converted to atmospheric transmission using $A(\lambda)= 10^{-k(\lambda)/2.5}$.
At airmasses other than 1, $A(\lambda)$ must be raised to the power $X$.

4. $r(\lambda)$ is the attenuation due to reflection from an aluminum surface,
taken from Allen (1973a). It is squared in eq.~2 because there are two such
reflections in the Cassegrain telescope systems used in this work.

5. $T(\lambda)$ is the transmission of the filter. For the work described here
and in Paper~II, virtually all of the observations were made at just two
telescopes, the 0.9-m reflectors at CTIO and KPNO, using the same $u$-band
filter.  Thus, detailed simulations of the $u$-band magnitudes have been
carried out for these specific telescope and camera systems.  The transmission
data for the $u$ filter are given in Appendix~A of the present paper.

6. Finally, $Q(\lambda)$ is the quantum efficiency of the CCD detector. Tables
for this were kindly provided by Walker (2000) for the Tek3 CCD used on the
CTIO 0.9-m telescope, and by Jacoby (1999) for the T2KA CCD used at the KPNO
0.9-m.

The calculations for the $B$ filter were handled slightly differently from the
above, since I wanted to simulate $B$ magnitudes that have been transformed to
the standard system (rather than instrumental magnitudes, which were calculated
for the $u$ band as just described).  Therefore, for $B$ I adopted the response
function tabulated by Azusienis \& Straizys (1969), choosing their
outside-atmosphere tabulation. (This table has also been re-published by Buser
\& Kurucz 1978, and widely used in simulations of the $UBVRI$ system.)  Since
Azusienis \& Straizys determined the full system response for the standard $B$
filter, I equated $r(\lambda)^2 \, T(\lambda) \, Q(\lambda)$ to their values in
my eq.~2.

In general, I did not re-calculate $B-V$ and $V-I$ colors, since these have
already been tabulated by LCB97 for their model atmospheres.  However, for the
red-leak simulations in \S5.3, I did need to calculate instrumental magnitudes
in the $I$ band.  For these computations, I used filter transmission curves
measured  by NOAO staff members; for the CTIO 0.9-m system, the functions were
obtained from the website http://www.ctio.noao.edu/instruments\slash filters/,
and for the KPNO 0.9-m from ftp:/\slash ftp.noao.edu\slash
kpno/filters/4Inch\_List.html.  

\subsection{Stellar Atmospheric Calibrations}

In order to simulate the behavior of the \textit{uBVI} system, in particular its
sensitivity to the basic stellar parameters, I have performed numerical
integrations using eq.~2 for model-atmosphere fluxes taken from LCB97.  For the
$B-V$ and $V-I$ colors, I simply adopted those tabulated by LCB97.  For
$u-B$, as described above, I calculated instrumental $u$ magnitudes for
the CTIO 0.9-m telescope with the Tek3 CCD camera, combined with $B$ magnitudes
calculated for the standard $B$ response function.  The calculations were done
for magnitudes outside the atmosphere (i.e., I set $X=0$ in eq.~2), and for
unreddened stars.  Magnitudes in the $u$ band were calculated for the main
ultraviolet bandpass alone, and excluded the contribution from the filter's red
leak (see below).

The constants in eq.~1 were set such that the $u-B$ color of the LCB97 model
atmosphere with $\Teff=9500$~K, $\log g=4.0$, and $\feh=-0.5$, which are very
close to the atmospheric parameters of Vega (Castelli \& Kurucz 1994), is
$u-B=1.0$.  The $B-V$ and $V-I$ colors of this model, tabulated by LCB97, are
of course essentially 0.0 (actually, they are $-0.012$ and $-0.021$,
respectively).  I have thus followed the precepts of \Str\ (1963), who set the
$u-b$ color of Vega and other A0~V stars to 1.0 in the $uvby$ system.  Given
the considerable drop in flux below the Balmer jump, this is a more realistic
choice than setting $u-B=0.0$, as was done, for example, in the classical $UBV$
system (Johnson \& Morgan 1953), or in other ``Vega-mag'' photometric systems.

Figure 4a shows the $u-B$ vs.\ $B-V$ color-color diagram for stars of solar
metallicity, effective temperatures of 5,000 to 14,000~K, and surface gravities
of $\log g=0.5$ to 4.5.  (Note that in this and similar diagrams, I plot $u-B$
increasing upwards, rather than downwards as is the convention in the $UBV$
system; this is done in order to have high-luminosity stars lie near the top of
the diagram, as is the convention in similar plots in the \Str\ system.)
Figure~4b is the same diagram, but for stars of $\feh=-2$.  Lines of constant
effective temperature (dashed) and of constant surface gravity (solid) are drawn
and labelled.  These figures show the high sensitivity of the $u-B$ color index
to gravity for stars of 5,000~K up to about 10,000~K\null.  As is well known,
above about 10,000~K, the Balmer jump (and hence the $u-B$ color index) is not
highly sensitive to $\log g$, but remains sensitive to $\Teff$\null. 
Unfortunately, however, both $u-B$ and $B-V$ are quite sensitive to metallicity,
as a comparison of Figures~4a and 4b shows, since both indices become
significantly bluer as $\feh$ is reduced.

The metallicity dependence can be mitigated to a considerable extent by adopting
a color difference analogous to the \Str\ $c_1$ index, which he defined as
$c_1=(u-v)-(v-b)$ (where $u$ here is, of course, \Str's $u$).  Such an index
retains a high sensitivity to gravity, but is less sensitive to metallicity (and
also to interstellar reddening). For the \textit{uBVI} system, I adopt the color
difference $(u-B)-(B-V)$, which I plot against $V-I$ in the color-color diagrams
in Figures~5a and 5b. In these figures, we see that $V-I$ is much less sensitive
to metallicity, as is $(u-B)-(B-V)$ except for stars below $\sim$6,000~K, where
some metallicity dependence remains.  As anticipated, the $(u-B)-(B-V)$ is very
sensitive to $\log g$. (It would, if desired, be possible to define more
complicated color-difference formulae with $V-I$ color terms,  with even less
sensitivity to metallicity and/or reddening, similar to the reddening-free
$[c_1]$ index used in the \Str\ system, e.g., \Str\ 1966.)

Detailed tables of the color grids are available upon request from the author.

\subsection{Zero-Age Main-Sequence Relation}

It is useful for many purposes to have color-color relations available for the
zero-age main sequence (ZAMS)\null. I have calculated such relations for $u-B$
vs.\ $B-V$ and $(u-B)-(B-V)$ vs.\ $V-I$, both at solar metallicity.  The
calculations were done by interpolation in $\Teff$ and $\log g$ in the
color-color grids, using the table of main-sequence surface gravities vs.\
spectral type given by Allen (1973b) and the table of effective temperatures
vs.\ spectral type given by Drilling \& Landolt (2000).

The ZAMS relations are given in Table~3. 

\section{Interstellar Extinction}

The dependence of the $u-B$ and $(u-B)-(B-V)$ color excesses upon interstellar
extinction was calculated using eq.~2 and varying the $B-V$ color excess, 
$E(B-V)$\null. As noted above, all of the calculations are based on the
reddening formula of Cardelli et al.\ (1989), with $R_V=3.1$. Because of the
finite width of the $u$ bandpass, the color-excess ratios depend weakly on the
stellar parameters, and are also slightly non-linear functions of $B-V$\null.
However, for most purposes, it will be adequate to adopt the ratios for a
lightly reddened, ``typical'' star. For a star of ($\Teff,\log
g,\feh)=(7000,2.5,0)$, i.e., a star lying near the middle of the color-color
grid of Figure~4a, and small amounts of reddening, the calculations show that
the color-excess ratios are given by:
\begin{eqnarray}
E(u-B) &=& 0.89 \, E(B-V) \, , \\
E[(u-B)-(B-V)] &=& -0.11 \, E(B-V) \, . \\
\noalign{For completeness, I note that the following relation has been given
for the color excess in $V-I$ by Dean, Warren, \& Cousins (1978):}\nonumber \\
E(V-I) &=& 1.25[1 + 0.06(B-V)_0 + 0.014E(B-V)] \, E(B-V)\, .
\end{eqnarray}

The slopes of these reddening vectors are plotted in Figures~4a-b and 5a-b. 

\section{Practical Considerations}

Other astronomers who may wish to implement the \textit{uBVI} system for their own
programs should take into account the following considerations when planning
and obtaining their observations and in reducing their data.

\subsection{Flat Fields}

It is usually satisfactory to perform flat-fielding of the CCD frames in $B$,
$V$, and $I$ by exposing on a uniformly illuminated white surface placed in
front of the telescope (``dome flats'').  However, in most dome-flat setups,
the illumination of the white surface is from incandescent lamps with a low
color temperature. These would be completely unsatisfactory for the $u$ filter,
because most of the signal would be transmitted through the filter's red leak. 
Hence it is essential that the $u$-band flats be obtained on the clear sky at
twilight.  These flats should generally be taken a few minutes after sunset
and/or a few minutes before sunrise, while the twilit sky is bright enough. 
With wide-field imaging systems, care should be taken to point the telescope so
as to avoid brightness gradients in the twilight illumination across the field
(see, for example, Chromey \& Hasselbacher 1996).

\subsection{Atmospheric Extinction}

Since the effective wavelength of the $u$ bandpass changes significantly with
stellar color (redder effective wavelength for redder stars) {\it and\/} it
changes with airmass (redder effective wavelength at higher airmass, due to the
steep increase in extinction at shorter wavelengths), it is desirable to
include both a color term and a non-linear airmass term in the extinction
corrections.

Simulations of the extinction behavior were performed using numerical
integrations based on eq.~2, varying both the stellar parameters and the
airmass $X$\null.  The system throughput of the CTIO 0.9-m camera and CCD was
adopted, and the extinction coefficients were fitted to the following equation:
\begin{equation}
u_{\rm instr}(X) = u_{\rm out} + k_1 X + k_2 X^2 \, ,
\end{equation}
where $u_{\rm instr}(X)$ is the instrumental $u$ magnitude measured at airmass
$X$, $u_{\rm out}$ is the $u$ magnitude that would be measured outside the
atmosphere, and $k_1$ and $k_2$ are the linear and quadratic extinction
coefficients. Least-squares fits to the simulated magnitudes show that $k_1$
can be represented adequately as a linear function of $B-V$ color,
\begin{equation}
k_1 = a + b(B-V) \, ,
\end{equation}
and that $k_2$ is essentially constant.

Eq.~6 therefore becomes
\begin{equation}
u_{\rm instr}(X) = u_{\rm out} + [a + b(B-V)] X + k_2 X^2 \, ,
\end{equation}
and the fit to the simulations yielded $a=0.615$, $b=-0.033$, and $k_2=-0.007$.

Extinction simulations were also run for the other three filters, and it was
found (as is generally adopted in $BVI$ photometry) that there are no
significant color or non-linear terms for the $V$ and $I$ filters, nor is there
a significant non-linear term for the $B$ extinction. However, a $b$ color term
of about $b=-0.03$ is appropriate for the $B$ filter.  

In practical observing situations, the recommended procedure is always to
observe a few standard fields at both low and high airmass during the night,
and to solve by least squares for the $a$ coefficient (which is indeed observed
to vary significantly from night to night).  If there is a sufficient range of
color among the standard stars, $b$ can also be solved for, or alternatively
simply taken to be $-0.033$.  However, in most actual situations, there will be
insufficient observations to solve for $k_2$, and it is recommended that it
simply be adopted as $k_2=-0.007$.  The best practice is to observe the
standard stars (apart from the extinction observations) and program stars over
as small a range of airmass as possible, so as to lessen the impact of
uncertainties in the extinction coefficients.

Note that most data-taking systems record the airmass at the start of the
exposure, not at the photon-weighted effective midpoint; this correction is
important in the $u$ band at high airmass and/or for long exposures, and the
IRAF routine \hbox{\it setairmass\/} should be used to calculate the effective
airmass.

\subsection{Red Leak}

As mentioned above, and shown in Figure~3, the $u$-filter glass combination
chosen for the \textit{uBVI} system has a significant red leak at about 7100~\AA\null.
Ideally, this leak should have been suppressed by adding another filter (such
as liquid or crystal CuSO$_4$), or by applying a leak-blocking coating to the
filter similar to that used for the SDSS $u'$ filter (Fukugita et~al.\ 1996). 
However, in the interests of economy, simplicity, and  making the throughput of
the $u$ filter as high as possible in the main bandpass, it was decided not to
attempt to suppress the red leak.

Simulations of the contribution of the red leak to the total photon count in
$u$ frames were calculated using eq.~2 for a range of stellar parameters and
airmasses. Two observational approaches were considered: (1)~calculate the red
leak as a fraction of the total signal in the $u$ band as a function of stellar
color and airmass, and (2)~take advantage of the fact that the red leak lies at
the short-wavelength edge of the $I$ bandpass, so that the red-leak signal in
the $u$ band should be a simple function of the $I$-band signal, with a weak
color term.

The simulations suggest that either approach is viable, but that the second one
(basing the red-leak correction on the measured photon count in the $I$ band)
has the advantage of no appreciable dependence on airmass.  Least-squares fits
to the simulations (for the CTIO 0.9-m system) yielded the following formulae:
\begin{eqnarray} 
\log({\rm RL}/u) &=& -2.662 + 2.019(B-V) -3.545(B-V)^2 \nonumber\\ 
&& + 3.336(B-V)^3 - 0.9086(B-V)^4 \nonumber\\ 
&& + [0.2076 - 0.001157(B-V) - 0.01902(B-V)^2](X-1.0) \\ 
\noalign{\noindent and}\nonumber \\
{\rm RL}/I &=& 0.0002268 -0.0000680(B-V) \nonumber\\ 
&& + 0.0000536(B-V)^2 -0.0000300(B-V)^3 \, , 
\end{eqnarray} 
where RL is the contribution to the $u$ counts due to the red leak, $u$ is the
total detected photon counts (main band plus red leak), $B-V$ is the color of
the star on the standard system, $I$ is the photon count in the $I$ band, and
$X$ is airmass.  All photon counts are per unit time.  It should be noted that
the corrections of eqs.~9 and~10 are to be made to the observed
inside-atmosphere $u$ counts {\it before\/} the atmospheric extinction is
removed.  Similar equations were calculated for the KPNO 0.9-m system (see
Paper~II).

Eq.~9 shows that, near the zenith, the red leak is predicted to contribute
about 1\% of the total $u$ signal for stars with $B-V=0.8$, 10\% of the
signal at $B-V=1.47$, and 30\% at $B-V=1.85$.  Thus, since the \textit{uBVI} system is intended primarily for
the study of stars of spectral types A through early G, the red leak can
usually simply be neglected.  For the most accurate work, the correction should
be made, using eq.~10 if $I$-band observations have been made, or eq.~9, if a
system similar to that of the CTIO 0.9-m telescope is used.  Alternatively, in
many cases it will be most practical (especially if the necessary laboratory
measurements of the camera and filter system are not available) to reduce the
$u$-band photometry to the system defined by the standard stars listed in
Paper~II, without making any explicit red-leak corrections. If the reddest
stars are excluded, any remaining adjustments for red leak will be absorbed in
linear and quadratic terms in $B-V$ in the transformation equation.  In the
standard stars listed in Paper~II, we have excluded all stars with $B-V>1.45$,
and have found that such transformations reproduce the red-leak-corrected $u$
magnitudes adequately.

\subsection{Reduction Procedures}

In summary, it is recommended that \textit{uBVI} photometric observations be 
reduced as follows:

1. Reduce the $BVI$ observations to the system defined by the Landolt (1992)
standard stars, using conventional techniques to remove atmospheric extinction
and then transform the instrumental magnitudes to the standard system.

2. If desired and if possible, correct each $u$ observation {\it inside the
atmosphere\/} for red leak, as described above (eq.~9 or 10).

3. Then, using standard fields observed at low and high airmass, solve for the
$u$-band extinction coefficients in eq.~8 (adopting the values recommended
above for $b$ and $k_2$).

4. After removing atmospheric extinction, fit the instrumental
outside-atmosphere magnitudes of the standard stars, $u_{\rm out}$, to the
standard values from Paper~II, $u_{\rm std}$, using a conventional transformation
equation of the form
\begin{equation}
u_{\rm out} = u_{\rm std} + c + d (B-V) + e (B-V)^2 \, ,
\end{equation}
where $u$ is the standard magnitude, and $B-V$ is the color index on the
standard system.   

5. An alternative to the above procedure is to perform both the extinction
correction and the transformation to the standard-star system through a single
least-squares fit to an equation of the following form:
\begin{equation}
u_{\rm instr} = u_{\rm std} + [a + b (B-V)] X + k_2 X^2 + c + d (B-V) + e
(B-V)^2 \, .
\end{equation}

In most practical cases, it will be adequate to set $k_2=0$.

6. Finally, using the above extinction and transformation equation(s), 
transform the program-star $u$ observations to the standard system.

\section{Conclusion}

I have described a CCD-based \textit{uBVI} photometric system that is highly optimized
for measurement of the Balmer jump in faint stars of spectral types A through
early G\null. Since the size of the Balmer jump is very sensitive to stellar
surface gravity in such stars, the \textit{uBVI} system should be useful in
applications involving determination of stellar luminosities and measurement of
extragalactic distances through high-luminosity standard candles of
Populations~I and~II\null. This system should also find application in the
study of phenomena on the horizontal branch in globular clusters (e.g.,
Grundahl et al.\ 1999), or indeed in a variety of settings involving stars too
faint for efficient observations in the \Str\ $uvby$ system.

Future papers in this series will report results of extensive \textit{uBVI}
observations of Galactic globular clusters and the halos and disks of nearby
galaxies.

\acknowledgments

The writer thanks Abi Saha and Tom Kinman for many useful discussions of
$u$-band photometry that guided the adoption of an optimal \textit{uBVI} system. STScI
postdocs Michael Siegel, Laura Fullton, and David Alves contributed much hard
work to the development of the system and data reductions. Ed Carder gave
useful practical advice about filters, constructed the $u$ filter used for the
standard-star work described in Paper~II, and provided a transmission curve for
it. Arlo Landolt and Peter Stetson provided much useful data and advice on
standard stars and the $UBVRI$ system.  Alistair Walker, George Jacoby, Mario
Hamuy, Masataka Fukugita, and Bruce Margon provided essential data tables. This
work was supported in part by the NASA UV, Visible, and Gravitational
Astrophysics Research and Analysis Program through grants NAG5-3912 and
NAG5-6821.

\appendix

\section{Details of the Thuan-Gunn \emph{u} Filter}

Following the precepts of Thuan \& Gunn (1976), the writer has had constructed
several $u$ filters of various sizes.  These filters are fabricated from Schott
glasses as follows: 4~mm UG11 + 1~mm BG38.  For comparison, the UG11 filter,
but with different thicknesses, is also used for \Str\ $u$ (8~mm UG11 + 1~mm
WG3) and SDSS $u'$ (1~mm UG11 + 1~mm BG38).

All of the standard-star observations described in Paper~II (Siegel \& Bond
2005) were accomplished with a $4\times4$ inch $u$ filter kindly constructed by
Mr.~Ed Carder of Kitt Peak National Observatory. Carder also kindly
measured the transmission curve for this filter using the Lambda 9
spectrophotometer at KPNO\null. The data are presented in Table~4.

\begin{figure}[h]
\begin{center}
\includegraphics[height=3.9in]{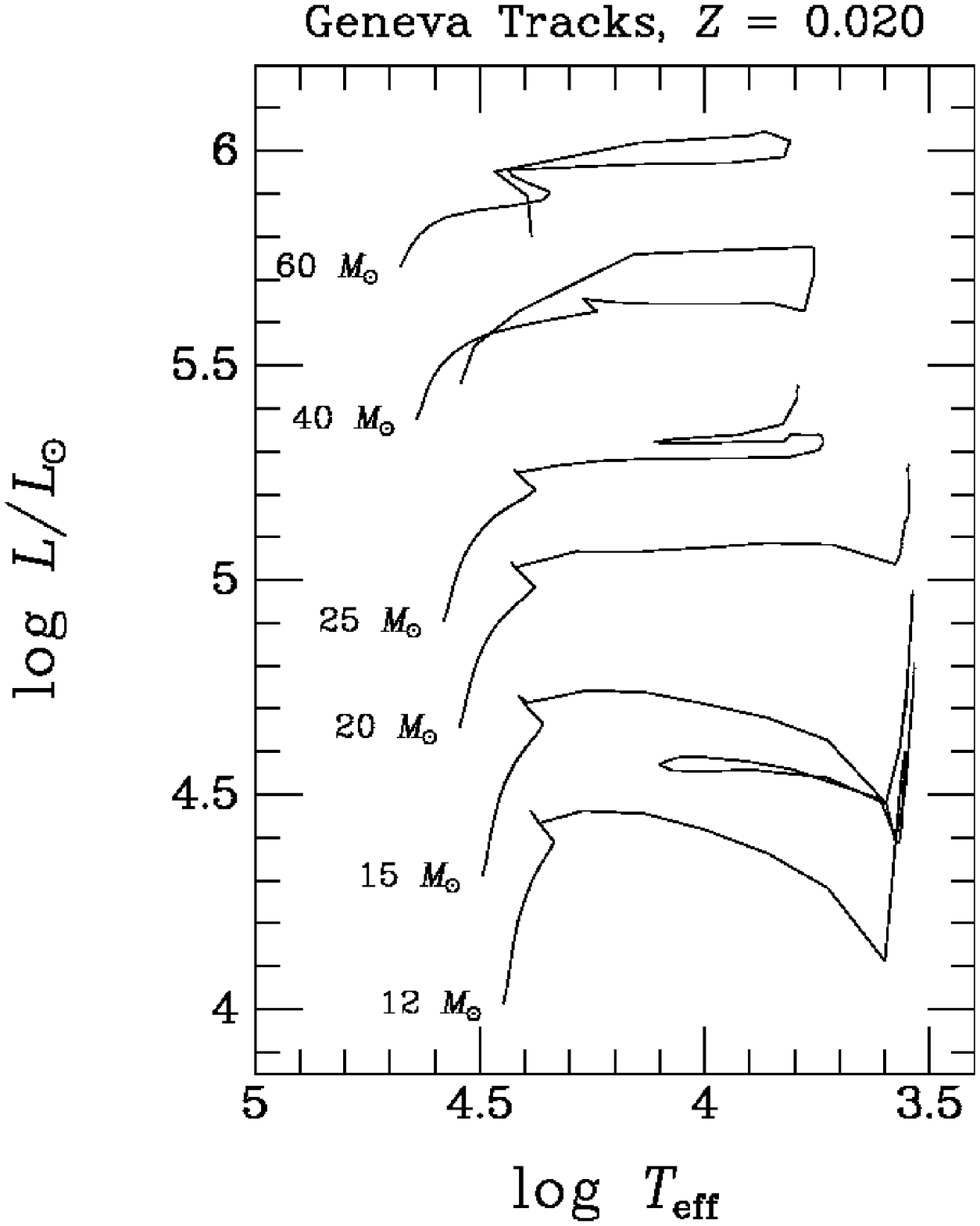}
\hfill
\includegraphics[height=3.9in]{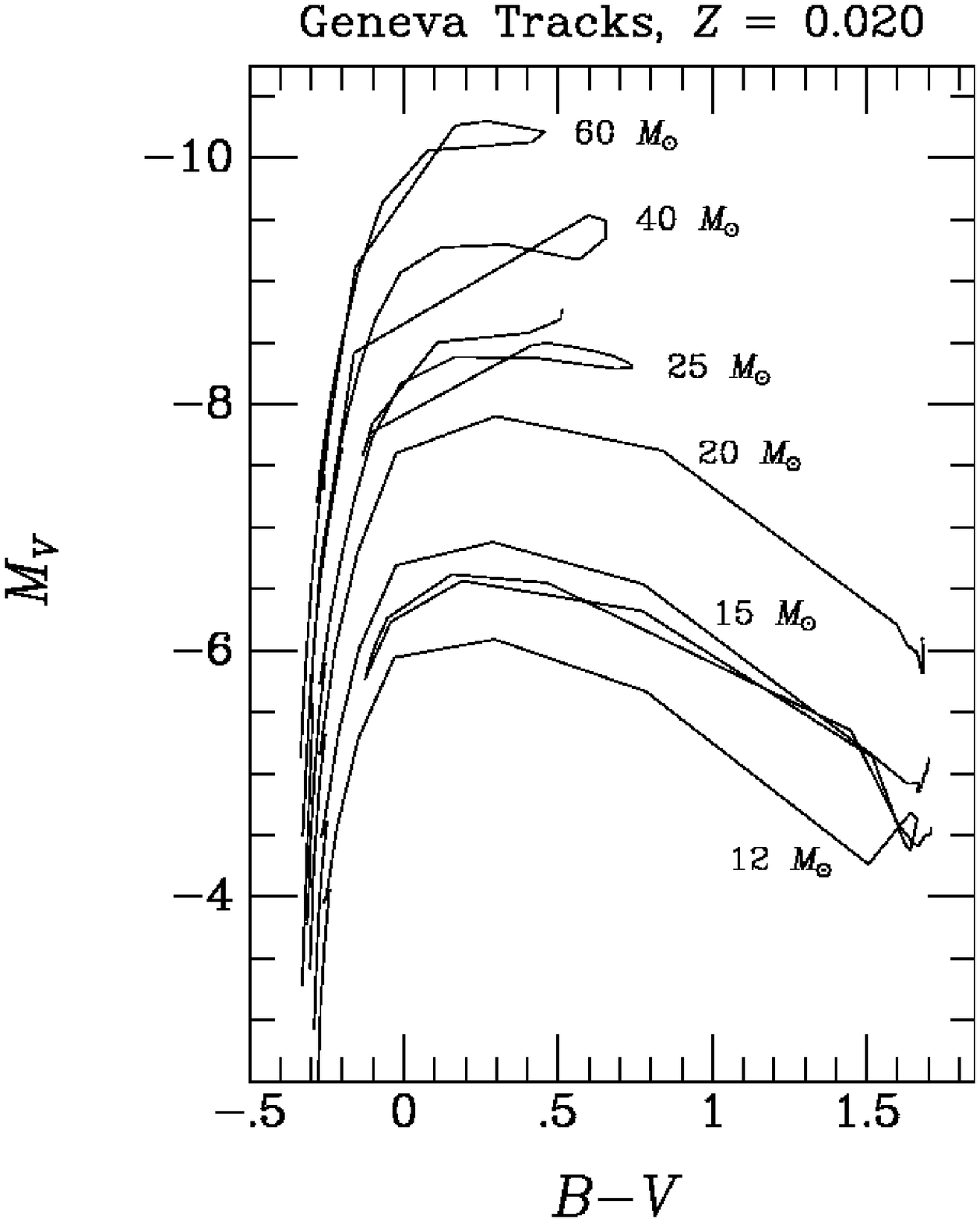}
\end{center}
\figcaption{{\bf Left:} Theoretical evolutionary tracks from the Geneva group
(Meynet et al.\ 1994) in the standard theoretical coordinates, for stars of
initial masses of 12--60~$M\subsun$. {\bf Right:} Same tracks transformed to
observational coordinates of absolute visual magnitude vs.\ $B-V$ color.  This
plot shows that yellow supergiants, with $B-V$ colors of about 0 to 0.5, are
{\it the visually brightest non-transient stars in galaxies}, because of the
behavior of the bolometric correction.}
\end{figure}

\begin{figure}
\begin{center}
\includegraphics[width=5in]{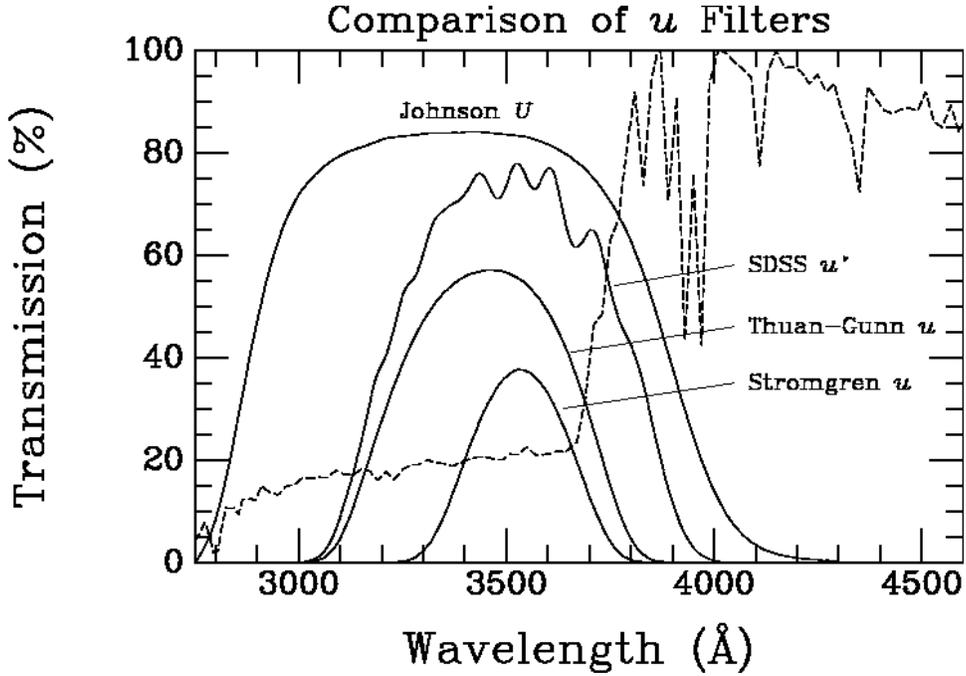}
\end{center}
\figcaption{Comparison of the transmission curves ({\it solid lines}) for four
candidate UV filters. Sources for the curves are given in the notes to Table~1.
Note that these curves do not include the atmospheric transmission function,
which will truncate the curves on the short-wavelength side, nor any
instrumental or detector throughput factors.  The {\it dashed line\/} plots the
flux curve of a model atmosphere with $(\Teff,\log g,\feh)=(7000,1,0)$ in order
to show the location of the Balmer jump. As discussed in the text and shown in
Table~1, the Thuan-Gunn $u$ filter offers the best combination of sensitivity
to the Balmer jump plus high throughput.}
\end{figure}

\begin{figure}
\begin{center}
\includegraphics[width=5in]{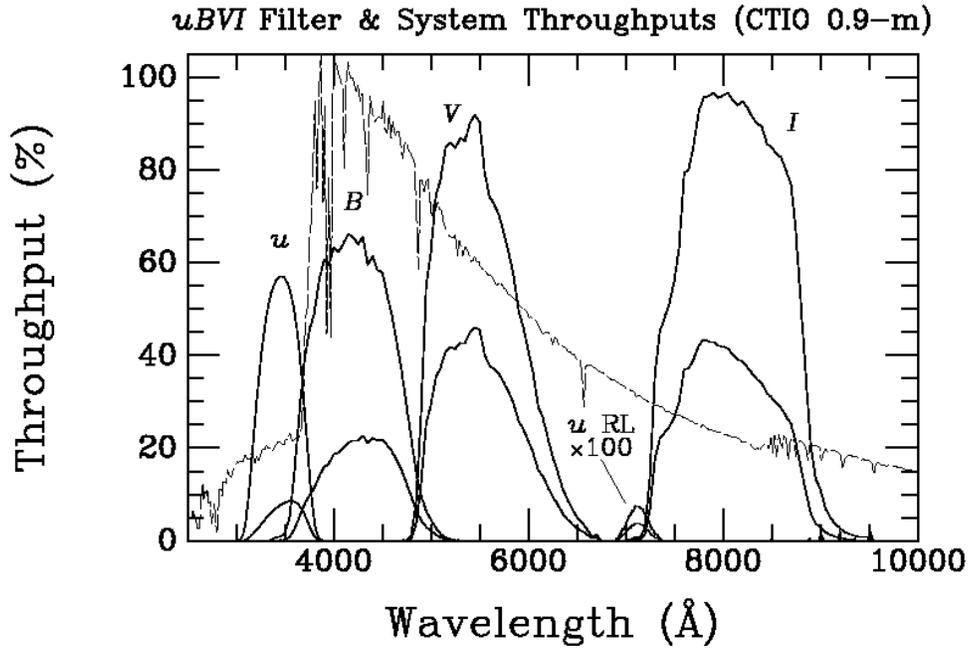}
\end{center}
\figcaption{Filter transmission curves for the \textit{uBVI} filters ({\it upper solid
lines}) and system throughput curves  for the Cerro Tololo 0.9-m telescope/CCD
camera combined with the atmospheric transmission at airmass 1.2 ({\it lower
solid lines}). The red-leak transmission and system throughput of the Thuan-Gunn
$u$ filter is also plotted, scaled up by a factor of 100. Also plotted is the
flux distribution for a model atmosphere with $(\Teff,\log g,\feh)=(7000,1,0)$
({\it light dashed curve}), taken from Lejeune et al.\ 1997.}
\end{figure}

\begin{figure}
\begin{center}
\includegraphics[width=4.5in]{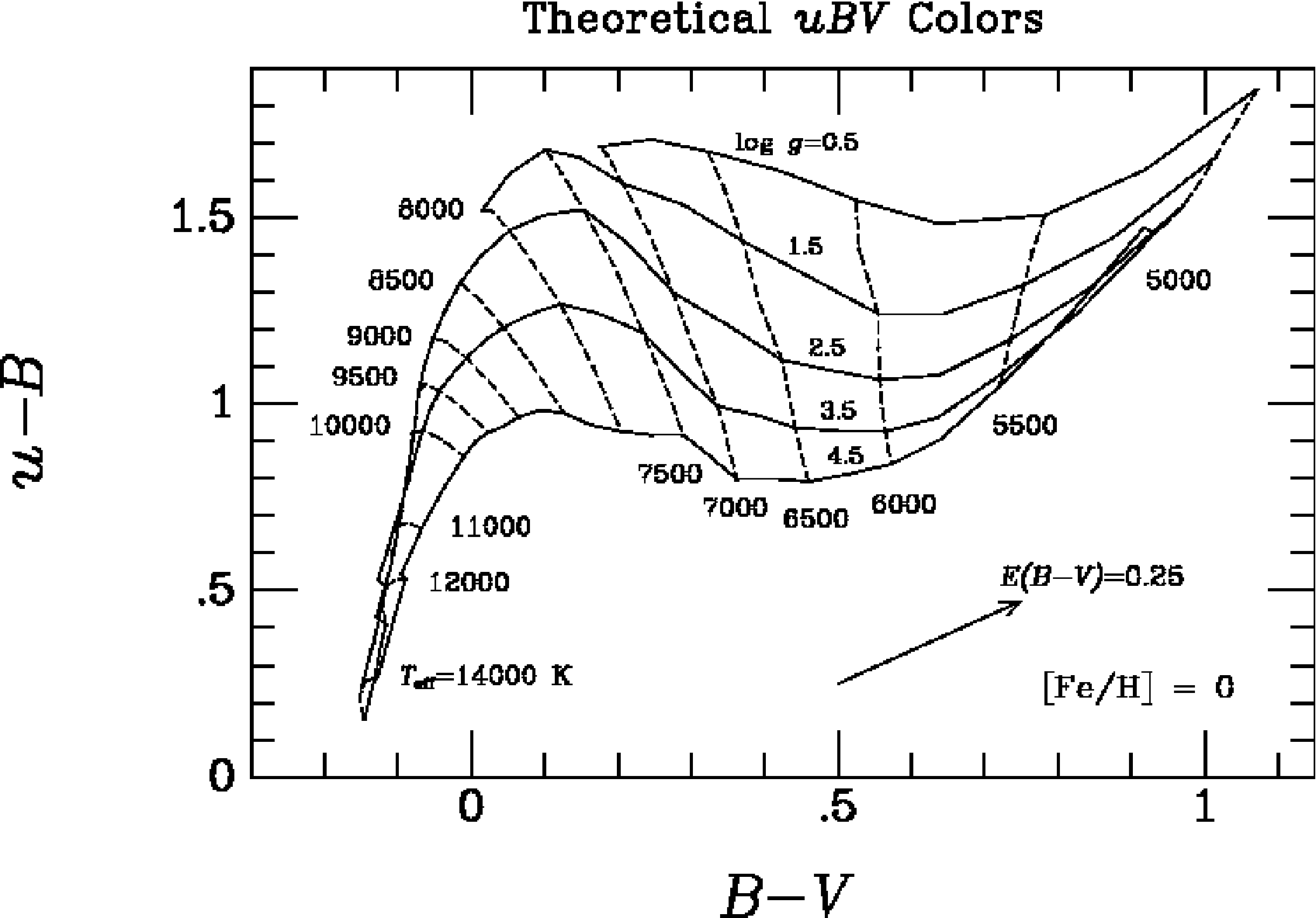}
\vskip .5in
\includegraphics[width=4.5in]{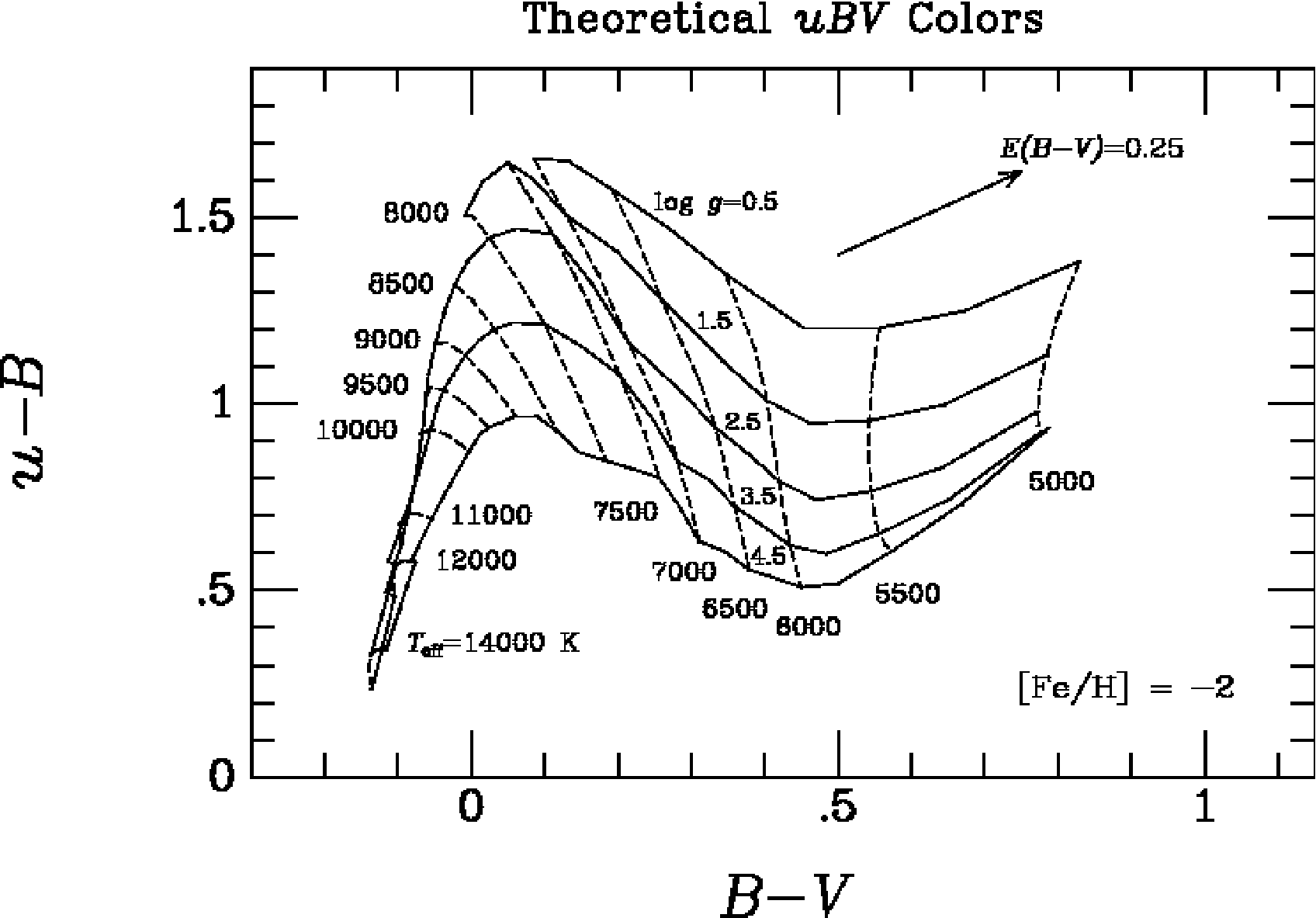}
\end{center}
\figcaption{$u-B$ vs.\ $B-V$ color-color relations for stars of solar
metallicity ({\bf a, top}) and 1\% of solar metallicity ({\bf b, bottom}). {\bf
Dashed lines} are lines of constant effective temperature, and {\bf solid
lines} are lines of constant surface gravity, as labelled. Slopes of reddening
vectors are indicated. Note that $u-B$ is very sensitive to $\log g$ for stars
up to $\Teff\simeq10,000$~K. $u-B$ and $B-V$ are, however, also sensitive to
metallicity.}
\end{figure}

\begin{figure}
\begin{center}
\includegraphics[width=4.5in]{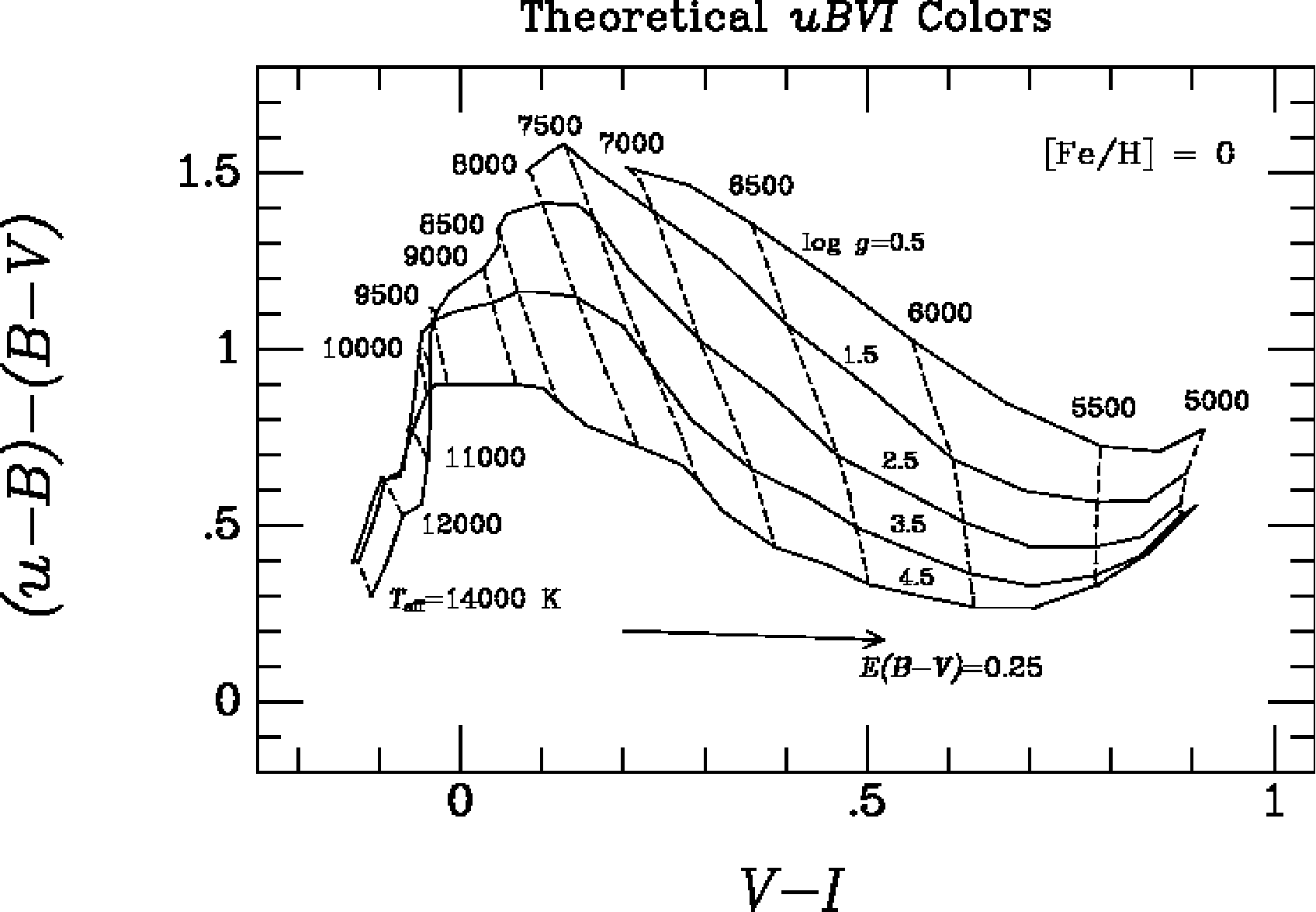}
\vskip .5in
\includegraphics[width=4.5in]{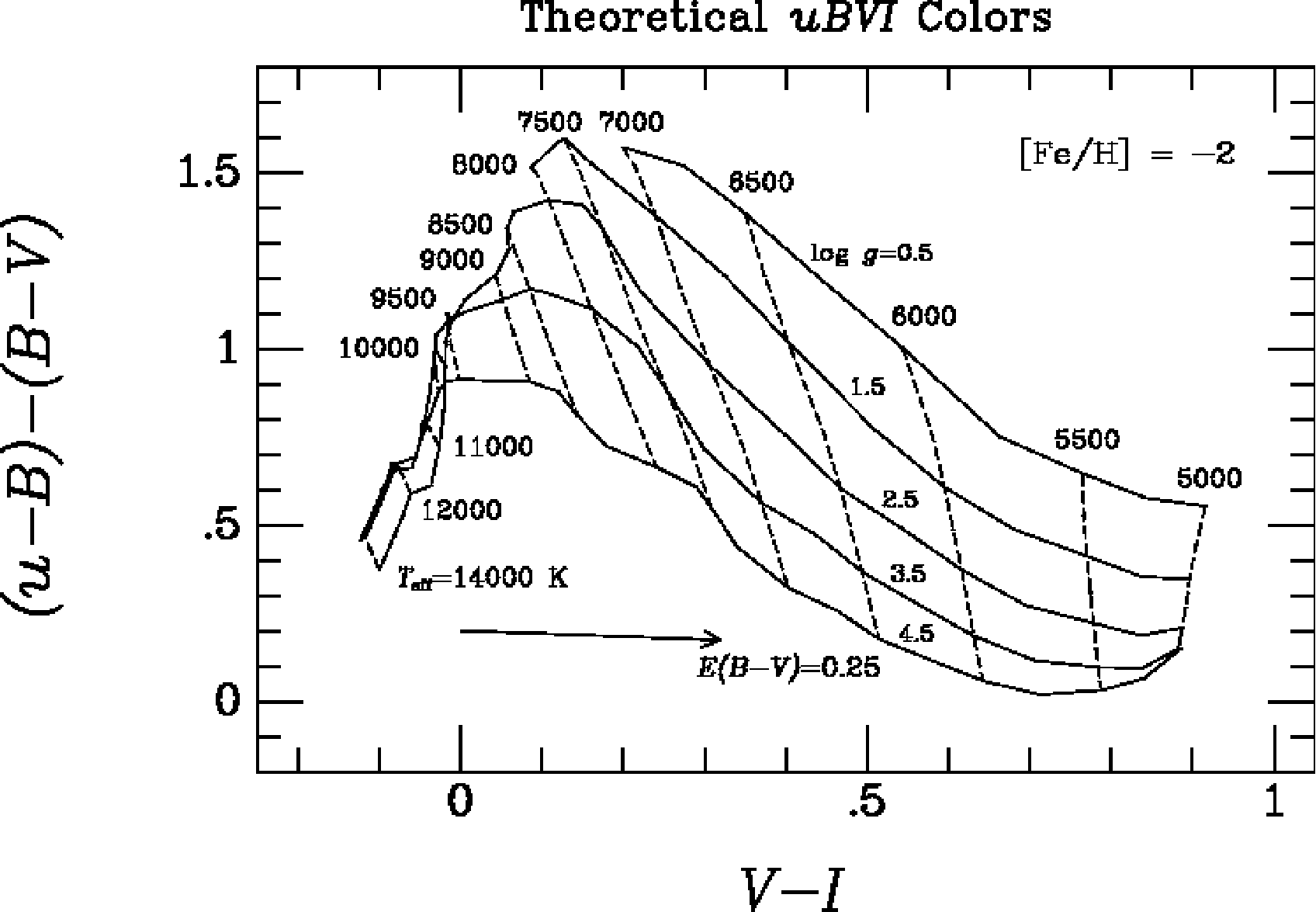}
\end{center}
\figcaption{$(u-B)-(B-V)$ vs.\ $V-I$ color-color relations for stars of solar
metallicity ({\bf a, top}) and 1\% of solar metallicity ({\bf b, bottom}).
Labelling of $\Teff$ and $\log g$ is as in Figures~4a-4b, and slopes of
reddening vectors are indicated. The $(u-B)-(B-V)$ color difference, which is
an analog of the $c_1$ index in the \Str\ $uvby$ system, has the advantage of
lower sensitivity to metallicity and reddening than the $u-B$ index, while
retaining high sensitivity to surface gravity. $V-I$ is also considerably less
sensitive to metallicity than $B-V$.}
\end{figure}

\clearpage

\begin{singlespace}

% the following turn "?" into a blank space, to make table column alignment
% nicer
\catcode`?=\active
\def?{\phantom{1}}

\begin{deluxetable}{lcc}
\tablecaption{Figures of Merit for Candidate UV Filters}
\tablewidth{0in}
\tablehead{
\colhead{Filter} & \colhead{$\Delta(u-B)$} & \colhead{Relative exposure time} }
\startdata
Thuan-Gunn $u$ &	 0.841 & 1.00 \\
\Str\ $u$      &	 0.889 & 2.01 \\
SDSS $u'$      &         0.488 & 1.36 \\
Johnson $U$    &	 0.380 & 1.44 \\
\enddata
\tablecomments{Col.~1: Filter name. Sources of filter transmission curves are
as follows. Thuan-Gunn $u$: Table~3 of this paper; SDSS $u'$: Fukugita 2004;
\Str\ $u$: ftp://ftp.noao.edu/kpno/filters/4indata/kp1538; Johnson $U$: Landolt
1992, Table~8; Johnson $B$: http://www.ctio.noao.edu/instruments/filters/;
col.~2: Change in instrumental $u-B$ color index at airmass 1.2 for 7000~K
stars when $\log g$ is changed from 4.5 to 1.0; col.~3: Relative exposure times
required to measure $\Delta(u-B)$ to a given percentage error. See text for
further explanation.}
\end{deluxetable}

\begin{deluxetable}{lcccl}
\tablecaption{Properties of \textit{uBVI} Filters}
\tablewidth{0in}
\tablehead{
\colhead{Filter} & \colhead{Vega mag} & \colhead{$\lambda_{\rm eff}$ (\AA)} & 
\colhead{FWHM (\AA)} & \colhead{Reference}  }
\startdata
$u$ & 1.00 & 3530 & ?400  & Thuan \& Gunn 1976     \\
$B$ & 0.00 & 4747 & 1409  & Fukugita et al.\ 1996  \\
$V$ & 0.00 & 5470 & ?826  & Fukugita et al.\ 1996  \\
$I$ & 0.00 & 8020 & 1543  & Fukugita et al.\ 1996  \\
\enddata
\end{deluxetable}

\begin{deluxetable}{lccccc}
\tablecaption{Zero-Age Main-Sequence Relation at $\feh=0$}
\tablewidth{0in}
\tablehead{\colhead{$\Teff$ (K)} & \colhead{$\log g$} &
\colhead{$B-V$} & \colhead{$u-B$} & \colhead{$V-I$} & 
\colhead{$(u-B)-(B-V)$} }
\startdata
\llap{1}4000 & 4.04 & \llap{$-$}0.140 &  0.261 &  \llap{$-$}0.130 &  0.401 \\
\llap{1}3000 & 4.04 & \llap{$-$}0.124 &  0.378 &  \llap{$-$}0.114 &  0.502 \\
\llap{1}2500 & 4.04 & \llap{$-$}0.115 &  0.448 &  \llap{$-$}0.106 &  0.563 \\
\llap{1}2000 & 4.04 & \llap{$-$}0.105 &  0.528 &  \llap{$-$}0.097 &  0.633 \\
\llap{1}1500 & 4.04 & \llap{$-$}0.112 &  0.547 &  \llap{$-$}0.074 &  0.659 \\
\llap{1}1000 & 4.06 & \llap{$-$}0.087 &  0.680 &  \llap{$-$}0.064 &  0.766 \\
\llap{1}0500 & 4.10 & \llap{$-$}0.063 &  0.786 &  \llap{$-$}0.055 &  0.850 \\
\llap{1}0000 & 4.13 & \llap{$-$}0.034 &  0.896 &  \llap{$-$}0.046 &  0.930 \\
9750 & 4.14 & \llap{$-$}0.017 &  0.945 &  \llap{$-$}0.038 &  0.962 \\
9500 & 4.17 & \llap{$-$}0.002 &  0.969 &  \llap{$-$}0.022 &  0.971 \\
9250 & 4.19 &	   0.018 &  0.988 &	       0.006 &  0.972 \\
9000 & 4.21 &	   0.044 &  1.014 &	       0.059 &  0.970 \\
8750 & 4.24 &	   0.073 &  1.037 &	       0.089 &  0.970 \\
8500 & 4.26 &	   0.105 &  1.036 &	       0.105 &  0.938 \\
8250 & 4.28 &	   0.142 &  1.019 &	       0.139 &  0.872 \\
8000 & 4.30 &	   0.189 &  0.996 &	       0.202 &  0.811 \\
7750 & 4.31 &	   0.235 &  0.968 &	       0.259 &  0.736 \\
7500 & 4.33 &	   0.282 &  0.954 &	       0.284 &  0.675 \\
7250 & 4.34 &	   0.319 &  0.895 &	       0.318 &  0.575 \\
7000 & 4.34 &	   0.358 &  0.827 &	       0.383 &  0.469 \\
6750 & 4.34 &	   0.408 &  0.825 &	       0.447 &  0.417 \\
6500 & 4.35 &	   0.457 &  0.811 &	       0.500 &  0.354 \\
6250 & 4.37 &	   0.513 &  0.827 &	       0.564 &  0.314 \\
6000 & 4.39 &	   0.572 &  0.848 &	       0.631 &  0.276 \\
5750 & 4.44 &	   0.639 &  0.907 &	       0.705 &  0.268 \\
5500 & 4.49 &	   0.722 &  1.052 &	       0.783 &  0.330 \\
5250 & 4.49 &	   0.812 &  1.231 &	       0.844 &  0.419 \\
5000 & 4.50 &	   0.918 &  1.474 &	       0.904 &  0.556 
\enddata
\end{deluxetable}

\begin{deluxetable}{cccc}
\tablecaption{Transmission Curve for Thuan-Gunn $u$ Filter}
\tablewidth{0in}
\tablehead{
\colhead{Wavelength (\AA)} & \colhead{Transmission (\%)} & 
\colhead{Wavelength (\AA)} & \colhead{Transmission (\%)} }
\startdata
   3000 &  ??.000 &	6900 &	   .000 \\
   3010 &  ??.015 &	6910 &	   .005 \\
   3020 &  ??.050 &	6920 &	   .005 \\
   3030 &  ??.115 &	6930 &	   .010 \\
   3040 &  ??.240 &	6940 &	   .015 \\
   3050 &  ??.465 &	6950 &	   .015 \\
   3060 &  ??.850 &	6960 &	   .020 \\
   3070 &  ?1.420 &	6970 &	   .025 \\
   3080 &  ?2.225 &	6980 &	   .025 \\
   3090 &  ?3.300 &	6990 &	   .030 \\
   3100 &  ?4.675 &	7000 &	   .035 \\
   3110 &  ?6.375 &	7010 &	   .040 \\
   3120 &  ?8.280 &	7020 &	   .045 \\
   3130 &   10.455 &	7030 &	   .050 \\
   3140 &   12.770 &	7040 &	   .055 \\
   3150 &   15.245 &	7050 &	   .055 \\
   3160 &   17.745 &	7060 &	   .060 \\
   3170 &   20.275 &	7070 &	   .065 \\
   3180 &   22.830 &	7080 &	   .065 \\
   3190 &   25.270 &	7090 &	   .070 \\
   3200 &   27.705 &	7100 &	   .075 \\
   3210 &   30.095 &	7110 &	   .075 \\
   3220 &   32.360 &	7120 &	   .075 \\
   3230 &   34.595 &	7130 &	   .075 \\
   3240 &   36.670 &	7140 &	   .075 \\
   3250 &   38.700 &	7150 &	   .070 \\
   3260 &   40.610 &	7160 &	   .070 \\
   3270 &   42.385 &	7170 &	   .070 \\
   3280 &   44.125 &	7180 &	   .065 \\
   3290 &   45.560 &	7190 &	   .060 \\
   3300 &   47.065 &	7200 &	   .055 \\
   3310 &   48.375 &	7210 &	   .050 \\
   3320 &   49.610 &	7220 &	   .045 \\
   3330 &   50.755 &	7230 &	   .045 \\
   3340 &   51.785 &	7240 &	   .040 \\
   3350 &   52.670 &	7250 &	   .035 \\
   3360 &   53.425 &	7260 &	   .030 \\
   3370 &   54.190 &	7270 &	   .025 \\
   3380 &   54.770 &	7280 &	   .025 \\
   3390 &   55.340 &	7290 &	   .020 \\
   3400 &   55.730 &	7300 &	   .015 \\
   3410 &   56.155 &	7310 &	   .015 \\
   3420 &   56.420 &	7320 &	   .015 \\
   3430 &   56.705 &	7330 &	   .010 \\
   3440 &   56.810 &	7340 &	   .010 \\
   3450 &   57.025 &	7350 &	   .005 \\
   3460 &   56.965 &	7360 &	   .005 \\
   3470 &   57.025 &	7370 &	   .000 \\
   3480 &   56.920 &	7380 &	   .000 \\
   3490 &   56.795 &	7390 &	   .000 \\
   3500 &   56.565 &	7400 &	   .000 \\
   3510 &   56.210			\\
   3520 &   55.840			\\
   3530 &   55.315			\\
   3540 &   54.800			\\
   3550 &   54.120			\\
   3560 &   53.270			\\
   3570 &   52.345			\\
   3580 &   51.300			\\
   3590 &   50.195			\\
   3600 &   48.910			\\
   3610 &   47.480			\\
   3620 &   45.990			\\
   3630 &   44.270			\\
   3640 &   42.455			\\
   3650 &   40.460			\\
   3660 &   38.305			\\
   3670 &   36.005			\\
   3680 &   33.675			\\
   3690 &   31.165			\\
   3700 &   28.555			\\
   3710 &   25.870			\\
   3720 &   23.165			\\
   3730 &   20.415			\\
   3740 &   17.715			\\
   3750 &   15.040			\\
   3760 &   12.565			\\
   3770 &   10.210			\\
   3780 &   ?8.080			\\
   3790 &   ?6.180			\\
   3800 &   ?4.570			\\
   3810 &   ?3.255			\\
   3820 &   ?2.235			\\
   3830 &   ?1.460			\\
   3840 &   ??.905			\\
   3850 &   ??.535			\\
   3860 &   ??.290			\\
   3870 &   ??.150			\\
   3880 &   ??.065			\\
   3890 &   ??.025			\\
   3900 &   ??.005			
\enddata
\end{deluxetable}
\end{singlespace}
\end{document}